\documentstyle[epsfig]{l-aa}
\input epsf
\newcommand{\nh}{N${\rm _H}~$}
\newcommand{\cgs}{$erg~cm^{-2}s^{-1}$}
\newcommand{\einstein}{{\it Einstein} }
\newcommand{\rosat}{{\it ROSAT} }
\newcommand{\sax}{{\it Beppo}SAX }
\newcommand{\exosat}{{\it EXOSAT} }
\newcommand{\etal}{{\it et al.$~$}}

\newcommand{\ergs}{{erg~cm$^{-2}$~s$^{-1}~$}}
\newcommand{\aox} {{$\alpha_{\rm ox}~$}}
\newcommand{\aro} {{$\alpha_{\rm ro}~$}}
\newcommand{\aoxaro}{{$\alpha_{\rm ox}-\alpha_{\rm ro}~$}}
\newcommand{\lsim}{{\lower.5ex\hbox{$\; \buildrel < \over \sim 
\;$}}}
\newcommand{\gsim}{{\lower.5ex\hbox{$\; \buildrel > \over \sim 
\;$}}}
\newcommand{\vovm}{{$V/V_{\rm m}$}~}
\newcommand{\vova}{{$V_{\rm e}/V_{\rm a}$}~}

\newcommand{\vovaave}{{$\langle V_{\rm e}/V_{\rm a} \rangle$}~}

\begin{document}

   \thesaurus{         
              (; 
               )} 
   \title{The BeppoSAX 2-10 keV Survey}


   \author{P. Giommi\inst{1}, M. Perri\inst{1} \& F. Fiore\inst{1,2,3}}

   \offprints{P. Giommi}

\institute{
   {\inst{1}\sax Science Data Center, ASI,
    Via Corcolle, 19,
    I-00131 Roma , Italy}\\
   {\inst{2}Osservatorio Astronomico di Roma,
   Via Frascati, 33,  
    I-00044 Monteporzio, Italy}\\
   {\inst{3}Harvard-Smithsonian Center for Astrophysics,
   60 Garden st. Cambridge, MA 02138 USA}\\
} 

   \date{Received  ; accepted }

   \maketitle
   \markboth{Giommi, Perri \& Fiore: The \sax 2-10 keV Survey}{}

%
%
%
%
%
%
\begin{abstract} 

We present the results of a 2-10 keV \sax survey based on  
140 high galactic latitude MECS fields, 12 of which are deep 
exposures of ``blank'' parts of the sky. 
The limiting sensitivity is $5\times10^{-14}$ \cgs where  
about 25\% of the Cosmic X-ray Background (CXB) is resolved 
into discrete sources. The logN-logS function, built with a 
statistically complete sample of 177 sources, is steep and in 
good agreement with the counts derived from ASCA surveys. A CXB 
fluctuation analysis allowed us to probe the logN-logS down to about 
$1.5\times10^{-14}$ \cgs where the contribution of discrete sources to 
the CXB grows to $\sim 40-50\%$. 

A hardness ratio analysis reveals the presence of a wide range of 
spectral shapes and that a fairly large fraction of sources appear 
to be heavily absorbed, some of which showing soft components. 

A comparison of the flux distribution of different subsamples confirms 
the existence of a spectral hardening with decreasing flux. This effect 
is probably due to an increasing percentage of absorbed sources 
at faint  fluxes, rather than to a gradual flattening of the spectral 
slope.  
Nearly all the sources for which adequate \rosat exposures exist, 
have been detected in the soft X-rays. This confirms that 
soft spectral components are present even in strongly absorbed objects, 
and that a large population of sources undetectable below a few keV 
does not exist. 

A \vova test provides evidence for the presence of cosmological 
evolution of a magnitude similar to that found in soft X-ray 
extragalactic sources. Evolution is present both in normal and 
absorbed sources, with the latter population possibly evolving faster,
although this effect could also be the result of complex 
selection effects. 

\end{abstract} 

 
\section{Introduction} 
 
Since the early days of X-ray astronomy many surveys have regularly 
addressed one of the most intriguing and most intensively studied 
issues in this field: the nature of the Cosmic X-ray Background (CXB) 
(e.g. Seward \etal 1967, Boldt, Desai \& Holt 1969, 
Schwartz, Murray \& Gursky 1976,  Giacconi \etal 1979, 
Marshall \etal 1980, Maccacaro \etal 1991, Giommi \etal 1991, 
Garmire \etal 1992, Hasinger et al 1993, 1998, McHardy \etal 1998, 
Zamorani \etal 1999). 
It is now widely accepted that at least a substantial part of the 
CXB (above 1 keV) is due to the combined emission of discrete 
extragalactic sources, mostly AGN, a large fraction of which could be
heavily obscured (Setti \& Woltjer 1989, Madau, Ghisellini \& Fabian 1994, 
Comastri \etal 1995, Comastri 1999, Fabian 1999, Gilli \etal 1999).

All surveys carried out with the first generation of X-ray 
telescopes were technically limited to the soft band where 
photoelectric absorption in the Galaxy, and within the emitters,
induces strong biases. 
As most of the energy of the CXB is instead located in the 
hard X-rays, despite the very important results obtained with 
\einstein and \rosat, some crucial questions still remain unanswered. 

Over the past few years, ASCA and \sax pushed the high energy 
limit of X-ray optics to about 10 keV removing some of the problems 
associated to photoelectric absorption. 
The analysis of ASCA data (Inoue \etal 1996, Georgantopoulos \etal 
1997, Cagnoni, Della Ceca and Maccacaro 1998, Ueda \etal 1998,1999, 
Della Ceca \etal 1999a) and the \sax  results (Giommi \etal 1998, 
Giommi, Fiore, \& Perri 1999,  Fiore \etal 1999, Fiore \etal 2000a) 
have indeed revealed that absorption plays a crucial role in the 
making of the CXB and, consequently, that optical surveys might have 
missed much of the accretion power in the Universe. 
These findings also showed that the picture is less simple than 
anticipated due to a) the presence of complex X-ray spectra with soft 
components even in heavily obscured objects (Giommi, Fiore \& Perri 1999,
Della Ceca \etal 1999a) , and b)
a range of optical properties wider than that found in optical 
surveys based on color selection.
(Schmidt \etal 1998, Fiore \etal 1999, La Franca \etal 2000, 
Fiore \etal 2000b, Lehmann \etal 2000).

The new generation of powerful X-ray mirrors  aboard Chandra and 
XMM-Newton 
have already started probing the CXB at very faint fluxes 
(Brandt \etal 2000, Fiore \etal 2000b, Mushotzky \etal 2000). Over 
the next several years these results, together with the outcome of  
massive optical identification campaigns necessitating  
the power of the largest existing optical telescopes, will definitively 
settle many of the open issues. Some issues, however, will 
probably have to wait for future hard X-ray telescopes operating 
in the 10-50 keV band, one of the last unexplored energy windows 
where the bulk of the CXB power is emitted.

In this paper we present the results of a 2-10 keV survey carried out 
with the MECS instruments aboard \sax (Boella et al. 1997a,b)
covering the flux range $\sim 1\times 10^{-14} - \sim 1\times 10^{-
12}$ \ergs. 
This flux interval is useful to address some of the still open issues, 
and is helpful for the proper assessment of the bright tail of 
the much deeper Chandra and XMM-Newton logN-logS. This may be 
necessary as the determination of the bright part of a logN-logS 
requires the analysis of large areas of sky. With XMM-Newton or 
Chandra this can only be achieved by searching for serendipitous 
sources in a large number of images, many of which will inevitably be 
centered on targets that may be as bright as the sources 
sought, lowering the probability of finding other bright serendipitous 
sources in the same field. This bias also effects the \sax survey but 
at higher fluxes, above $\sim 1\times 10^{-12}$ \ergs.

The results of a parallel \sax survey, carried out in the harder 
5-10 keV band (the HELLAS Survey) are reported in a separate paper 
(Fiore \etal 2000a). The HELLAS survey was specifically designed to
address the issue of hard sources and to take advantage of the MECS 
PSF, which is sharper in the high energy band. 
 
\section{The data} 

The data used for the survey presented here include 140 MECS fields, 
17 of which (12 deep exposures and 5 less exposed) are centered 
on ``blank'' parts of the sky. 
The remaining fields are from observations made as part of one of the 
main \sax programs, and have been taken from the \sax Science Data 
Center (SDC) public archive (Giommi \& Fiore 1997) after the 
expiration 
of the proprietary period. Most of the deep pointings are secondary 
Narrow 
Fields Instruments (NFI) observations, that is exposures pointing 
90 degrees away from a primary Wide Field Camera (WFC, Jager \etal 
1997) target (usually the Galactic Center) and centered on a position 
chosen to optimize the satellite roll angle.  
In a few cases the deep exposures were centered near Polaris, 
the \sax default pointing position in case of safe mode. 

The public fields have been selected according to the following 
criteria: 
\begin{enumerate}
\item the pointing is at high Galactic latitude ($|b|>20$)
\item the exposure is longer than 10,000 seconds; 
\item the field was public before the end of December 1999;
\item the target of the observation is not too bright (count rate $<$ 0.2 
cts/s) or extended. 
\item the field does not include regions such as SMC, LMC, M31, M33 
etc..
\item when two fields partially or totally overlap, the one with
the deepest exposure is chosen.
\end{enumerate}

 The sample so defined covers $\sim$ 45 square degrees of high 
galactic latitude sky.

Although the number of MECS fields is similar (140) to that of the 
HELLAS survey (142) the overlap between the two surveys is not very large. 
This is mainly due to two factors: 1) a relatively large number of fields 
could not be used in the 2-10 keV band because of condition (4) 
above, which is necessary to avoid 
contamination problems connected to the 2-10 keV PSF which is 
significantly wider than that in the 5-10 keV band, and 2) we include 
fields that have become public as late as December 1999, whereas the 
HELLAS survey only uses fields that were public before April 1999.
Other differences come from the fact 
that the 2-10 keV survey must avoid the area obscured by the MECS 
beryllium support window (which is instead transparent above 5 keV,
Boella \etal 1997b) and that 
the exclusion area around the target is 6 arcminutes radius in the 2-10 
keV survey and 4 arcminutes in HELLAS. Finally the useful field of 
view has been chosen to be 24 arcminutes in the 2-10 keV band and 
25 arcminutes in HELLAS.

\subsection{Data analysis} 

The procedure followed for the data analysis is very similar to that 
used for the \sax  HELLAS survey, which is described in detail in Fiore 
\etal 2000a. 
We will not repeat all the details here; we will instead summarize 
the main points and describe the differences with respect to the 
HELLAS survey.

From the MECS cleaned and calibrated event files we have built X-ray 
images taking photons with energy channels (PI) between 44 and 220 
corresponding to 2-10 keV. Data from all three MECS units were
co-added in sky coordinates for observations carried out before the 
failure of MECS1 on May 7 1997; the sum of MECS2 and MECS3 were 
used in all other cases.
All images have been searched for sources using the detect routine of 
the  XIMAGE package (Giommi \etal 1991) modified as described in 
Giommi \etal 1998 and in Fiore \etal 2000a. 
A statistical probability threshold of $5 ~ 10^{-4}$ of being a 
fluctuation of the local background was chosen so that very few 
spurious sources should be included in the sample. 
All the fields centered on ``blank'' parts of the sky (i.e. not on  
a previously known X-ray source) were searched over 
the full field of view (up to an off-axis angle of 24 arcminutes), 
avoiding the window support structure and two circular regions (8.7 
arcmin radius) near the edge of the 
field of view and centered on the on-board calibration sources. 
All the observations taken from the public archive have been 
searched for serendipitous sources outside the central 6 arcminutes to 
exclude the region immediately surrounding the target of the 
observation. 

Finally each candidate detection has been carefully inspected in 
various energy bands to remove spurious detections near the edge of 
the field of view or close to the on-board calibration sources.  

\subsection{The Sky Coverage} 

The sensitivity of the MECS instrument, besides the obvious 
dependence on 
exposure time, is a complex and strong function of the position in the 
field of view. Consequently the area covered by our survey at any 
given flux (usually known as the sky coverage) is a complex function 
of flux.

Two basic factors are responsible for this dependence: (1) the effective 
area decreases at large off-axis angles (vignetting effect) and (2) 
the Point Spread Function (PSF) degrades with distance from the 
center. Both these effects depend on energy. 

The minimum detectable count rate ($cr_{min}$) in our MECS images 
can be analytically described as follows:
\begin{equation}
cr_{min} = cr_o / \sqrt{t} ~~ (1+0.0077 R^{1.88})
\end{equation}	
where $cr_o$ is the minimum source count rate detectable at the 
center of the field of view, $t$ is the exposure time in seconds, and R is 
the off-axis radius expressed in arcminutes. 
The dependence on $t$ is given by $1 / \sqrt{t}$ as our X-ray 
images are always background limited since the minimum exposure 
time
considered is $ 10,000 s$.
The normalization $cr_o$ has been derived by comparing the 
predictions of  equation (1) to the source count rates extracted from 
our database. 
We adopted the values $cr_o = 0.24$ and $cr_o = 0.20 $ for three and 
two MECS units respectively. These values are somewhat conservative 
since some real sources but not necessarily all can still be 
detected just below the threshold. For the purposes of this paper, however, 
we prefer to stay somewhat above the ultimate MECS sensitivity limit 
by rejecting all the sources below the count rate given by eq (1). 
As our simulations show (see section 4) this procedure ensures that a) 
the number of spurious sources is reduced to a very 
small percentage; b) the problem of source confusion is minimized; and 
c) a uniform detection capability over the entire field of view is achieved.  

The MECS detectors are very stable, both during single orbits, and over 
long time periods. The total background level is due to instrumental 
noise and to the cosmic signal. The first component has been 
monitored during periods when the sky is occulted by the Earth and 
has been found to be only slightly decreasing with time with a total 
change of a few percent since the beginning of the mission. We 
conclude 
that, to a good approximation, the values of $cr_o$ can be considered 
constant throughout the mission. 

The sky coverage of our survey has been computed applying the 
sensitivity law of eq. (1) to all the 140 MECS fields, taking 
into account that the areas behind the window support structure and 
around the calibration sources were not used. 
Count rates have been converted to 2-10 keV 
flux assuming a power law spectral model absorbed by an amount of 
\nh equal to the Galactic value as determined by the 21 cm 
measurements of Dickey \& Lockman (1990).
The MECS absolute flux calibration has been checked through several 
observations of the Crab Nebula carried out at regular intervals 
throughout the mission. No variations in the MECS sensitivity
have been detected (Sacco 1999).

Figure 1 shows the sky coverage of our survey for 
three different power law energy slopes, $\alpha =0.2, 0.6$ and 
$1.0$. Although some dependence on the spectral slope 
is clearly present, this is not very strong, and the differences 
are no larger than 20-30\% for slopes as different as those 
considered.

\begin{figure} 
\epsfig{figure=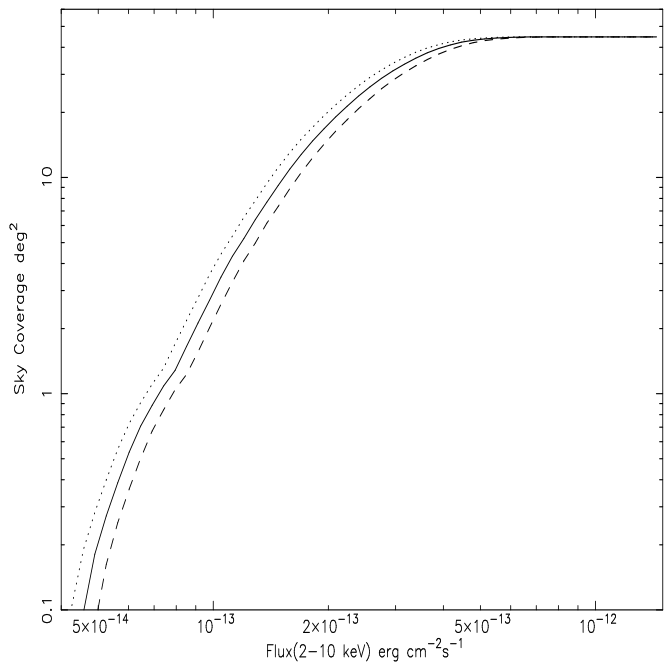,height=10.0cm,width=10.cm,angle=-90} 
\caption{The sky coverage of the 2-10 keV \sax Survey for 
three power law energy spectral slopes: $\alpha =0.2$, dashed line,
$\alpha =0.6$, solid line, and  $\alpha =1.0$, dotted line.}
\label{fig2} 
\end{figure} 

\subsection{A statistically well defined sample of 2-10 keV sources} 

The sources included in our 2-10 keV survey consist of all detections 
with count rate above the threshold given by eq (1). 
The complete lists, comprising 177 sources,  
is presented in table 1 where column 1 gives the source name; column 
2 gives other designations in case the source was previously known; columns 
3 and 4 give the Right Ascension and Declination; column 5 gives the 
2-10 keV flux calculated assuming a power law spectrum with energy 
index $\alpha $ = 0.7; column 6 gives the source classification, and column 7 
give the redshift of extragalactic sources when available.  

The uncertainty in the source positions is generally of the order of 
1 arcminute, although at large off-axis angles, and in a few cases 
where some source confusion might be present, this error can be as 
large as 1.5-2.0 arcminutes (for a more detailed 
discussion about \sax position uncertainties see Fiore \etal 2000a). 

Three quarters of the sources in the list are unidentified. The only 
identifications are from cross-correlations with catalogs of known 
objects, the NED and SIMBAD on-line systems, or from the published 
optical identifications of the HELLAS survey for common sources. 
The radius used for the correlation is 2 arcminutes; in case the 
candidate counterpart is found at a distance larger than 1.2 arcminutes 
we consider the association as tentative and we add a question mark 
in column 2 of table 1.
Of the 45 identified sources, 35 are AGN, 2 are late type stars, one 
is a RSCVn star system, and 7 are 
clusters of galaxies. We do not attempt to distinguish between type 1
and type 2 AGN since the identification process is inevitably biased in 
favor of type 1 AGN which are much better represented in 
astronomical catalogs than type 2 objects. 

From column 2 in table 1 we see that 96 \sax sources have soft X-ray 
counterparts in 
\einstein, \exosat or \rosat images. Whenever a source in our survey 
is within one of the \rosat public fields we have checked for 
possible soft X-ray counterparts. We have found that nearly all \sax 2-10 
keV sources are detected also in the 0.2-2.0 keV band if a reasonably 
deep image exist. 

The selection criteria listed in section 2 and equation (1) ensure that
our sample is statistically well defined and is suitable for statistical 
analysis. 

\newpage
\setcounter{table}{0}
\begin{table*}
\begin{minipage}{180mm}
\caption{List of X-ray sources.}
\begin{tabular}{llccccc}
\hline \hline
Source name& Other designation(s) & RA & Dec & Flux(2-10keV) & Class & redshift\\
(1SAXJ)   &                      & (J2000.0) &(J2000.0) & $erg~cm^{-2}~s^{-1}$&\\
\hline
0026.5~-1944 & &00 26 33.4 & ~-19 44 07.0 &   3.94$\times10^{-13}$ & unclassified & \\
0056.3~-2236 & 1WGA J0056.2-2235? &00 56 20.6 & ~-22 36 51.9 &   3.31$\times10^{-13}$ & unclassified & \\
0057.0~-2231 & 1WGA J0057.0-2231 &00 57 03.0 & ~-22 31 19.2 &   1.85$\times10^{-13}$ & unclassified & \\
0057.4~-2209 & 1WGA J0057.4-2210 &00 57 28.5 & ~-22 09 58.2 &   2.98$\times10^{-13}$ & unclassified & \\
0153.8+8910 & &01 53 51.4 & ~~89 10 13.6 &   8.28$\times10^{-14}$ & unclassified & \\
0135.9~-2954 & &01 35 54.8 & ~-29 54 45.8 &   1.11$\times10^{-13}$ & unclassified&  \\
0149.1+1353 & &01 49 06.9 & ~~13 53 39.9 &   4.03$\times10^{-13}$ & unclassified & \\
0200.5+0045 & 1WGA J0200.5+0046 &02 00 34.3 & ~~00 45 08.4 &   3.87$\times10^{-13}$ & unclassified  \\
0223.4~-0852 & 1WGA J0223.4-0852 &02 23 25.3 & ~-08 52 25.2 &   3.90$\times10^{-13}$ & unclassified&  \\
0223.9~-0843 & &02 23 57.0 & ~-08 43 50.2 &   1.17$\times10^{-13}$ & unclassified&  \\
0243.1~-0023 & 1WGA J0243.1-0025?&02 43 06.7 & ~-00 23 40.4 &   6.57$\times10^{-13}$ & HBL BL Lac? & \\
0249.6+3124 & &02 49 37.0 & ~~31 24 53.8 &   3.88$\times10^{-13}$ & unclassified & \\
0250.2+3110 & &02 50 12.4 & ~~31 10 40.8 &   2.91$\times10^{-13}$ & unclassified & \\
0313.5~-5509 & 1WGA J0313.4-5509 &03 13 30.6 & ~-55 09 48.7 &   1.58$\times10^{-13}$ & AGN & 1.378$^{(a)}$\\
0314.9~-5519 & 1WGA J0314.9-5519 &03 14 55.6 & ~-55 19 12.8 &   8.34$\times10^{-14}$ & Radio Galaxy & 0.387$^{(a)}$\\
0315.7~-5517 & 1WGA J0315.8-5517 &03 15 42.1 & ~-55 17 35.0 &   7.90$\times10^{-14}$ & AGN & 0.808$^{(a)}$\\
0315.7~-5528 & 1WGA J0315.7-5528 &03 15 44.9 & ~-55 28 24.6 &   4.83$\times10^{-13}$ & AGN & 0.464$^{(a)}$\\
0317.4~-5519 & 1WGA J0317.5-5519 &03 17 28.4 & ~-55 19 41.3 &   8.78$\times10^{-13}$ & AGN & 0.406$^{(a)}$\\
0325.3+0213 &  &03 25 18.4 & ~~02 13 59.2 &   5.24$\times10^{-13}$ & unclassified & \\
0336.8~-3615 & 1WGA J0336.9-3616 &03 36 51.2 & ~-36 15 34.6 &   4.99$\times10^{-13}$ & Blazar& 1.537$^{(b)}$\\
0406.1~-7100 & &04 06 06.3 & ~-71 00 44.4 &   6.53$\times10^{-13}$ & unclassified & \\
0406.7~-7115 & MS 0407.2-7123&04 06 45.0 & ~-71 15 52.8 &   4.66$\times10^{-13}$ & Cluster of gal. & 0.229$^{(c)}$ \\
 & 1WGA J0406.6-7116 & & & & &  \\
0407.5~-1217 & 1WGA J0407.5-1217&04 07 35.9 & ~-12 17 03.6 &   2.14$\times10^{-13}$ & unclassified & \\
0407.8~-7127 & EXO 0408.4-7134?&04 07 53.6 & ~-71 27 54.5 &   1.65$\times10^{-13}$ & M star &\\
0410.6~-7122 & 1RXP J041038-7123.8?&04 10 36.9 & ~-71 22 32.7 &   9.09$\times10^{-14}$ & unclassified & \\
0414.3~-5557 & &04 14 18.8 & ~-55 57 02.1 &   3.87$\times10^{-13}$ & unclassified & \\
0424.2~-5657 & &04 24 14.6 & ~-56 57 58.3 &   4.30$\times10^{-13}$ & unclassified & \\
0437.1~-4730 &1WGA J0437.1-4732 &04 37 10.2 & ~-47 30 40.9 &   4.25$\times10^{-13}$ & unclassified & \\
0438.7~-4727 &1WGA J0438.7-4727 &04 38 45.7 & ~-47 27 26.2 &   4.80$\times10^{-13}$ & Cand Blazar & \\
0515.2+0108 & 1WGA J0515.2+0109 &05 15 14.0 & ~~01 08 54.8 &   2.16$\times10^{-13}$ & unclassified & \\
0522.3~-3625 &1WGA J0522.2-3624 & 05 22 20.3&-36 25 03.4&  1.39$\times10^{-13}$& Cluster of gal. & 0.530$^{(d)}$\\
0523.6~-3630 &1WGA J0523.7-3630?&05 23 36.3 & ~-36 30 03.9 &   1.33$\times10^{-13}$ & unclassified & \\
0524.2~-3620 &1WGA J0524.2-3621&05 24 16.8 & ~-36 20 48.7 &   9.54$\times10^{-13}$ & unclassified  &\\
0536.8~-4402 &2E0535.2-4404 &05 36 48.8 & ~-44 02 00.7 &   4.71$\times10^{-13}$ & unclassified & \\
0538.8~-4413 & &05 38 50.1 & ~-44 13 35.8 &   1.30$\times10^{-13}$ & unclassified & \\
0539.8~-4357 & &05 39 52.1 & ~-43 57 30.1 &   2.75$\times10^{-13}$ & unclassified & \\
0548.6~-6052 & &05 48 37.9 & ~-60 52 17.0 &   5.76$\times10^{-13}$ & unclassified & \\
0549.9~-6123 & &05 49 58.8 & ~-61 23 18.0 &   3.28$\times10^{-13}$ & unclassified & \\
0549.9~-6102 & &05 49 59.5 & ~-61 02 48.7 &   1.08$\times10^{-13}$ & unclassified & \\
0550.6~-6058 & &05 50 41.8 & ~-60 58 57.2 &   1.17$\times10^{-13}$ & unclassified & \\
0613.2+7053 &1RXP J061321+7054.4 &06 13 14.1 & ~~70 53 47.0 &   1.10$\times10^{-13}$ & unclassified & \\
0613.8~-6054 & &06 13 52.6 & ~-60 54 10.2 &   8.74$\times10^{-14}$ & unclassified & \\
0613.9~-6100 & &06 13 59.1 & ~-61 00 52.4 &   4.92$\times10^{-14}$ & unclassified & \\
0623.7~-6913 & &06 23 43.6 & ~-69 13 50.6 &   1.19$\times10^{-13}$ & unclassified & \\
0625.4~-6918 & &06 25 27.9 & ~-69 18 30.2 &   1.61$\times10^{-13}$ & unclassified & \\
0718.9+7124 &1WGA J0718.9+7124&07 18 58.2&~~71 24 49.7&1.72$\times10^{-13}$& AGN Radio L. & 1.419$^{(e)}$\\
0720.7+7109 &1WGA J0720.8+7108 &07 20 42.5 & ~~71 09 45.0 & 1.31$\times10^{-13}$ & Cluster of gal. &\\
                 &HST J072049+71089 &           &              &                      &                 &    \\
0733.2+3204 & B2 0730+32 &07 33 17.7 & ~~32 04 43.4 & 3.23$\times10^{-13}$ & Double radio s.& \\
0733.3+3151 & &07 33 21.7 & ~~31 51 41.4 &   2.93$\times10^{-13}$ & unclassified  &\\
0733.9+3143 &1RXP J073401+3143.1&07 33 59.6& ~~31 43 27.4&1.78$\times10^{-13}$& unclassified & \\
\hline
\end{tabular}
\end{minipage}
\end{table*}
\newpage
\setcounter{table}{0}
\begin{table*}
\begin{minipage}{180mm}
\caption{List of X-ray sources (continued).}
\begin{tabular}{llccccc}
\hline \hline
Source name& Other designation(s) & RA & Dec & Flux(2-10keV) & Class & redshift\\
(1SAXJ)    &                      & (J2000.0) &(J2000.0) & $erg~cm^{-2}~s^{-1}$&\\
\hline
0741.9+7427 &1WGA J0742.0+7426 &07 41 57.9 & ~~74 27 15.2 &   4.13$\times10^{-13}$ & unclassified & \\
0743.2+7430 & MS 0737.0+7436 &07 43 13.1 & ~~74 30 12.6 & 1.12$\times10^{-12}$ & AGN & 0.312$^{(c)}$\\
& 1WGA J0743.1+7429 & & & & & \\
0837.6+2548 &1WGA J0837.6+2547 &08 37 36.6 & ~~25 48 03.4 &   4.10$\times10^{-13}$ & unclassified & \\
0838.7+2600 &1WGA J0838.8+2600?&08 38 47.5 & ~~26 00 25.7 &   2.50$\times10^{-13}$ & unclassified & \\
0838.9+2608 &1WGA J0838.9+2608 &08 38 59.4 & ~~26 08 03.7 &   1.95$\times10^{-12}$ & unclassified & \\
0914.8+4100 & &09 14 51.1 & ~~41 00 49.5 &   2.67$\times10^{-13}$ & unclassified & \\
0915.9+2928 &1WGA J0915.9+292?&09 15 56.9 & ~~29 28 34.4 &   2.50$\times10^{-13}$ & unclassified & \\
0916.3+2940 &1WGA J0916.3+2939?&09 16 19.6 & ~~29 40 09.8 &   3.48$\times10^{-13}$ & Cand Blazar & \\
0959.2~-2255 & 1RXS J095917.9-22550 &09 59 17.5 & ~-22 55 18.8 &   2.92$\times10^{-13}$ & unclassified & \\
1006.5~-2007 & 1WGA J1006.5-2005? &10 06 32.8 & ~-20 07 04.7 &   3.80$\times10^{-13}$ & unclassified & \\
1007.5~-2025 & 1WGA J1007.6-2025&10 07 33.1 & ~-20 25 23.3 &   1.09$\times10^{-13}$ & unclassified & \\
1007.6~-2012 & 1WGA J1007.6-2012&10 07 38.0 & ~-20 12 27.7 &   1.91$\times10^{-13}$ & unclassified & \\
1014.2+7318 & 1RXH J101420.5+731725 &10 14 13.9 & ~~73 18 54.6 &   2.21$\times10^{-13}$ & unclassified & \\
1015.9+7137 & &10 15 56.8 & ~~71 37 34.9 &   1.65$\times10^{-13}$ & unclassified & \\
1017.7+7113 & &10 17 46.1 & ~~71 13 14.9 &   1.97$\times10^{-13}$ & unclassified & \\
1017.9+7133 &  1WGA J1017.9+7133 &10 17 58.5 & ~~71 33 53.6 &   2.82$\times10^{-13}$ & unclassified & \\
1019.1+7131 & &10 19 06.3 & ~~71 31 47.7 &   9.54$\times10^{-14}$ & unclassified & \\
1020.3+1957 &1RXP J102013+1958.4? &10 20 20.0 & ~~19 57 08.5 &   1.48$\times10^{-13}$ & unclassified & \\
1020.8+1955 & MS 1018.2+2010 &10 20 53.2 & ~~19 55 40.8 & 5.92$\times10^{-13}$ & AGN& 0.250$^{(c)}$\\
1021.5+7115 & &10 21 32.1 & ~~71 15 54.4 &   3.68$\times10^{-13}$ & unclassified & \\
1033.5+6854 & &10 33 34.3 & ~~68 54 35.1 &   1.45$\times10^{-13}$ & unclassified & \\
1036.2+5710 & &10 36 14.7 & ~~57 10 06.4 &   1.86$\times10^{-13}$ & unclassified & \\
1038.6+5714 & 1WGA J1038.7+5713? &10 38 37.8 & ~~57 14 19.7 &   1.38$\times10^{-13}$ & unclassified & \\
1050.1+3404 & 1WGA J1050.2+3404 &10 50 10.8 & ~~34 04 21.6 &   4.74$\times10^{-13}$ & unclassified & \\
1052.5+5723 & 1WGA J1052.6+5724 &10 52 33.4 & ~~57 23 57.4 &   9.03$\times10^{-14}$ & AGN & 1.113$^{(f)}$ \\
1052.8+5731 & 1WGA J1052.6+5731 &10 52 48.0 & ~~57 31 06.0 &   9.21$\times10^{-14}$ & AGN & n.a.$^{(g)}$ \\
1053.5+5725 & 1WGA J1053.5+5725 &10 53 32.7 & ~~57 25 39.0 & 1.18$\times10^{-13}$ & AGN & 0.784$^{(f)}$ \\
1054.3+5725 & 1WGA J1054.3+5725 &10 54 18.7 & ~~57 25 33.2 &   4.07$\times10^{-13}$ & AGN & 0.205$^{(f)}$ \\
1055.7+6028 & DM UMa&10 55 46.4 & ~~60 28 16.6 &   3.10$\times10^{-12}$ & RS CVn & \\
1057.6+5625 & &10 57 40.2 & ~~56 25 39.4 &   1.42$\times10^{-13}$ & unclassified & \\
1057.8+6022 & &10 57 48.3 & ~~60 22 15.1 &   1.61$\times10^{-13}$ & unclassified & \\
1106.9~-1815 & &11 06 56.9 & ~-18 15 40.0 &   1.27$\times10^{-13}$ & unclassified & \\
1106.9~-1801 & &11 06 57.5 & ~-18 01 26.3 &   4.67$\times10^{-13}$ & unclassified & \\
1107.2~-1838 & &11 07 14.9 & ~-18 38 27.2 &   4.85$\times10^{-13}$ & unclassified & \\
1120.3+1254 & 1WGA J1120.3+1253&11 20 21.8 & ~~12 54 08.1 &2.21$\times10^{-13}$ &unclassified &\\
1120.5+1306 & 1WGA J1120.6+1306 &11 20 35.5 & ~~13 06 52.8 & 1.72$\times10^{-13}$ & unclassified & \\
1134.9+7029& & 11 34 54.5 & ~~70 29 17.7& 4.89$\times10^{-13}$& unclassified & \\
1138.9~-1336 & &11 38 59.6 & ~-13 36 49.6 &   2.18$\times10^{-13}$ & unclassified & \\
1140.7~-1400 &PKS B1138-137 &11 40 43.1 & ~-14 00 56.1 & 1.03$\times10^{-12}$ & AGN Radio L. & n.a. \\
1156.9+6527& &11 56 59.9&~~65 27 48.2 & 2.03$\times10^{-13}$& unclassified & \\
1204.0+2808 &MS 1201.5+2824 &12 04 01.8 & ~~28 08 16.9 &   9.78$\times10^{-13}$ & Cluster of gal. &0.167$^{(c)}$\\
1217.6+4729 &ZW 1215+4745?? &12 17 41.7 & ~~47 29 44.6 &   5.91$\times10^{-13}$ & Cluster of gal.? & \\
1218.8+2958 &1WGA J1218.9+2959 &12 18 53.2 & ~~29 58 44.0 &   7.54$\times10^{-13}$ & AGN & 0.176$^{(h)}$\\
1218.9+3012 &1WGA J1218.8+3011? &12 18 55.2 & ~~30 12 46.4 &   2.81$\times10^{-13}$ & unclassified & \\
1219.7+4721 & NGC~4258 &12 19 45.8 & ~~47 21 09.8 & 1.32$\times10^{-13}$ & AGN & 0.654$^{(i)}$\\
& 1WGA J1219.8+4720 & & & & & \\
1221.6+2806 &1WGA J1221.5+2806 &12 21 36.7 & ~~28 06 45.0 &   3.34$\times10^{-13}$ & unclassified & \\
1221.7+7526 &MS 1219.9+7542  &12 21 45.9 & ~~75 26 19.7 &   4.12$\times10^{-13}$ & Cluster of gal. & 0.240$^{(c)}$\\
1222.2+2821 &1WGA J1222.2+2821 &12 22 12.8 & ~~28 21 12.9 &   5.43$\times10^{-13}$ & AGN& 0.028$^{(e)}$\\
1240.7~-3653 & &12 40 44.7 & ~-36 53 16.3 &   1.29$\times10^{-13}$ & unclassified & \\
1241.5+3251 &1WGA J1241.5+3250&12 41 32.5&~~32 51 42.4& 6.58$\times10^{-13}$& HBL BL Lac?&\\
\hline
\end{tabular}
\end{minipage}
\end{table*}
\newpage
\setcounter{table}{0}
\begin{table*}
\begin{minipage}{180mm}
\caption{List of X-ray sources (continued).}
\begin{tabular}{llccccc}
\hline \hline
Source name& Other designation(s) & RA & Dec & Flux(2-10keV) & Class & redshift\\
 (1SAXJ)   &                      & (J2000.0) &(J2000.0) & $erg~cm^{-2}~s^{-1}$&\\
\hline
1256.3+5909 & &12 56 20.8 & ~~59 09 20.4 &   2.21$\times10^{-13}$ & unclassified & \\
1305.2~-1021 & &13 05 13.1 & ~-10 21 01.5 &   2.28$\times10^{-13}$ & unclassified & \\
1305.5~-1032 & 1WGA J1305.5-1033 &13 05 32.4 & ~-10 32 46.3 & 3.13$\times10^{-12}$ & Blazar & 0.278$^{(j)}$\\
 & PKS~1302-102 & & & & &\\
1316.4+2901 & 1WGA J1316.4+2900 &13 16 24.8 & ~~29 01 13.3 &   1.85$\times10^{-13}$ & unclassified  \\
1316.7+2913 &1WGA J1316.8+2912 &13 16 47.5 & ~~29 13 42.2 &   1.08$\times10^{-13}$ & unclassified & \\
1321.5~-1654 & &13 21 31.6 & ~-16 54 16.8 &   2.92$\times10^{-13}$ & unclassified & \\
1321.7~-1635 & &13 21 47.5 & ~-16 35 41.1 &   5.53$\times10^{-13}$ & HBL BL Lac?& \\
1323.4~-1657 &IRAS F13207-1642? &13 23 24.1 & ~-16 57 18.9 &   1.79$\times10^{-12}$ & unclassified & \\
1338.8+0454 & &13 38 49.6 & ~~04 54 02.2 &   4.84$\times10^{-13}$ & unclassified & \\
1343.0+0001 & 1WGA J1342.9+0000 &13 43 01.3 & ~~00 01 35.4 &  3.97$\times10^{-13}$ & AGN & 0.804$^{(k)}$ \\
1343.9~-0008 & &13 43 57.4 & ~-00 08 52.7 &   8.03$\times10^{-14}$ & unclassified & \\
1344.9~-0016 & 1WGA J1344.9-0016 &13 44 58.0 & ~-00 16 27.8 & 6.85$\times10^{-13}$ & AGN & 0.245$^{(l)}$ \\
 & LBQS 1342-0000 & & & & &\\
1353.9+1820 &1RXP J135354+1819.9 &13 53 54.3 & ~~18 20 32.5 &   5.99$\times10^{-13}$ & AGN & 0.217$^{(h)}$ \\
1404.6+2613 &1RXP J140432+2611.9&14 04 36.1&~~26 13 09.3& 4.20$\times10^{-13}$& AGN & 0.585$^{(c)}$\\
1404.8+2553 &1WGA J1404.9+2552? &14 04 52.5 & ~~25 53 21.2 &   1.75$\times10^{-13}$ & unclassified & \\
1429.3+4451 & &14 29 21.6 & ~~44 51 40.1 &   1.75$\times10^{-13}$ & unclassified & \\
1430.7+4507 & &14 30 45.5 & ~~45 07 49.0 &   7.34$\times10^{-13}$ & unclassified & \\
1512.4~-0913 &1WGA J1512.4-0914? &15 12 29.8 & ~-09 13 21.5 &   1.22$\times10^{-13}$ & unclassified & \\
1514.4+3637 & MS 1512.4+3647& 15 14 25.0&~~36 37 57.1&5.43$\times10^{-13}$ & Cluster of gal. & 0.372$^{(c)}$ \\
1515.0+3658 &1WGA J1515.0+3657 &15 15 04.4 & ~~36 58 08.8 &   2.63$\times10^{-13}$ & AGN & 0.253$^{(m)}$ \\
 & CRSS J1515.0+3657 & & & & &\\
1528.1+1954 & &15 28 07.6 & ~~19 54 26.3 &   1.40$\times10^{-13}$ & unclassified & \\
1528.7+1945 & RX J152845+1944.5&15 28 45.4 & ~~19 45 04.2 & 3.85$\times10^{-13}$ &AGN?& 0.636$^{(n)}$ \\
1528.8+1938 & &15 28 50.6 & ~~19 38 11.5 &   2.76$\times10^{-13}$ & AGN & 0.657$^{(o)}$\\
1613.7+3412 & 1WGA J1613.6+3412 & 16 13 43.1 & ~~34 12 21.1 &   1.01$\times10^{-12}$ & Blazar & 1.401$^{(p)}$\\
 & 3EG J1614+3424 & & & & &\\
1633.6+5950 & &16 33 37.5 & ~~59 50 36.0 &   7.31$\times10^{-14}$ & unclassified & \\
1633.6+5942 & &16 33 37.9 & ~~59 42 22.3 &   5.37$\times10^{-14}$ & unclassified & \\
1634.1+5938 & &16 34 08.1 & ~~59 38 13.9 &   1.40$\times10^{-13}$ & unclassified & \\
1634.1+5946 & &16 34 08.6 & ~~59 46 23.7 &   1.83$\times10^{-13}$ & AGN & 0.341$^{(h)}$ \\
1635.5+5955 & 87GB 163448.4+600114 &16 35 31.7 & ~~59 55 13.7 &   2.83$\times10^{-13}$ & Cand Blazar & \\
1651.8+0441 & 1WGA J1651.8+0439? &16 51 49.7 & ~~04 41 04.2 & 1.02$\times10^{-12}$ & unclassified & \\
1653.8+0208 & &16 53 51.0 & ~~02 08 05.9 &   3.35$\times10^{-13}$ & unclassified & \\
1729.4+4904 &  RX J1729.5+4904 &17 29 26.4 & ~~49 04 22.9 &  1.32$\times10^{-13}$ & AGN & 0.960$^{(q)}$\\
1731.1+6054 & &17 31 10.4 & ~~60 54 57.1 &   6.95$\times10^{-14}$ & unclassified & \\
1741.2+6753 & 1RXS J174110.6+67521? &17 41 17.0 & ~~67 53 04.3 &   1.85$\times10^{-13}$ & unclassified & \\
1741.5+6805 & &17 41 33.3 & ~~68 05 02.3 &   1.53$\times10^{-13}$ & unclassified & \\
1744.1+6221 & &17 44 11.3 & ~~62 21 04.4 &   1.99$\times10^{-13}$ & unclassified & \\
1745.3+6242 & 1WGA J1745.4+6240? &17 45 21.8 & ~~62 42 09.1 & 1.84$\times10^{-13}$ & unclassified & \\
1751.5+6100 & &17 51 34.3 & ~~61 00 59.6 &   1.05$\times10^{-13}$ & unclassified & \\
1751.6+6119 & 1RXS J175140.6+61185? &17 51 36.8 & ~~61 19 56.3 &   1.04$\times10^{-13}$ & unclassified & \\
1752.6+6105 & 1RXS J175248.7+61050? &17 52 36.5 & ~~61 05 33.0 &   8.83$\times10^{-14}$ & unclassified & \\
1753.0+6120 & &17 53 03.3 & ~~61 20 31.3 &   1.29$\times10^{-13}$ & unclassified  & \\
1753.8+6059 & 1RXS J175347.9+60590 &17 53 50.9 & ~~60 59 36.7 &   1.65$\times10^{-13}$ & unclassified & \\
1757.7+6110 & &17 57 47.2 & ~~61 10 34.1 &   6.01$\times10^{-14}$ & unclassified & \\
1758.2+6118 & IRAS17579+6118?&17 58 14.9 & ~~61 18 40.8 &  6.70$\times10^{-14}$ & unclassified & \\
1759.1+7832 & 1RXP J175904+7833.4&17 59 07.9 & ~~78 32 43.2 &   1.34$\times10^{-13}$ & unclassified & \\
\hline
\end{tabular}
\end{minipage}
\end{table*}
\newpage
\setcounter{table}{0}
\begin{table*}
\begin{minipage}{180mm}
\caption{List of X-ray sources (continued).}
\begin{tabular}{llccccc}
\hline \hline
Source name& Other designation(s) & RA & Dec & Flux(2-10keV) & Class & redshift\\
(1SAXJ)   &                      & (J2000.0) &(J2000.0) & $erg~cm^{-2}~s^{-1}$&\\
\hline
1803.8+6110 & &18 03 53.1 & ~~61 10 24.3 &   7.08$\times10^{-14}$ & unclassified & \\
1804.4+6937 & 1WGA J1804.5+6937 &18 04 28.3 & ~~69 37 11.8 & 4.51$\times10^{-13}$ & AGN & 0.604$^{(e)}$\\
1805.3+6057 & 1RXS J180513.4+60563? &18 05 18.5 & ~~60 57 49.0 &   1.89$\times10^{-13}$ & unclassified & \\
1808.1+6947 & 1WGA J1808.1+6948 &18 08 08.4 & ~~69 47 44.2 &   4.60$\times10^{-13}$ & AGN & 0.096$^{(e)}$ \\
1818.9+6114 & 1RXS J181853.9+61162? &18 18 59.7 & ~~61 14 48.2 &   2.94$\times10^{-13}$ & unclassified & \\
1819.3+6056 & 1RXS J181922.4+60550? &18 19 18.5 & ~~60 56 24.1 &   2.63$\times10^{-13}$ & unclassified & \\
1819.6+6053 & &18 19 39.5 & ~~60 53 23.9 &   1.39$\times10^{-13}$ & unclassified & \\
1834.5+5147 & 1WGA J1834.4+5148? &18 34 34.4 & ~~51 47 43.2 &   2.05$\times10^{-13}$ & M star & \\
1936.0~-5238 & &19 36 05.4 & ~-52 38 04.1 &   2.29$\times10^{-13}$ & unclassified & \\
2040.5~-3222 & [WHO91] 2037-325 &20 40 33.2 & ~-32 22 22.5 &2.23$\times10^{-13}$ & AGN & 0.7$^{(r)}$\\
2041.1~-3247 & &20 41 09.0 & ~-32 47 01.1 &   5.94$\times10^{-13}$ & unclassified & \\
2041.9~-3215 & &20 41 59.8 & ~-32 15 36.0 &   1.77$\times10^{-13}$ & unclassified & \\
2055.7~-0417 & PKS J2055-0416 &20 55 46.9 & ~-04 17 07.6 & 4.03$\times10^{-13}$ & Blazar & 1.176$^{(p)}$\\
2055.8~-0452 & MS 2053.2-0503&20 55 49.7 & ~-04 52 09.6 &   3.81$\times10^{-13}$ & AGN & 0.281$^{(c)}$ \\
2056.2~-0446 & &20 56 12.6 & ~-04 46 11.3 &   1.03$\times10^{-13}$ & unclassified & \\
2225.2+2102 & 1WGA J2225.2+2102 &22 25 14.0 & ~~21 02 05.4 & 5.75$\times10^{-13}$ & unclassified & \\
2226.4+2111 &1WGA J2226.5+2111 &22 26 28.8 & ~~21 11 30.4 & 4.31$\times10^{-13}$&  AGN & 0.261$^{(c)}$ \\
2242.7+2935 & 1WGA J2242.7+2934 &22 42 46.8 & ~~29 35 15.7 &   2.59$\times10^{-13}$ & unclassified & \\
2244.7~-1213 & &22 44 42.9 & ~-12 13 20.2 &   8.78$\times10^{-13}$ & unclassified & \\
2245.5~-1200 & &22 45 35.8 & ~-12 00 27.4 &   2.48$\times10^{-13}$ & unclassified & \\
2305.9+0902 & &23 05 55.3 & ~~09 02 33.2 &   4.65$\times10^{-13}$ & Cand Blazar & \\
2306.9+0902 & &23 06 58.8 & ~~09 02 49.3 &   2.51$\times10^{-13}$ & unclassified & \\
2307.0+0850 & &23 07 01.0 & ~~08 50 04.5 &   2.01$\times10^{-13}$ & unclassified & \\
2327.4+0845 & 1WGA J2327.4+0845&23 27 26.1 & ~~08 45 49.8 &   9.73$\times10^{-14}$ & unclassified & \\
2327.5+0848 & 1WGA J2327.4+0849 &23 27 31.2 & ~~08 48 58.4 & 1.25$\times10^{-13}$ & unclassified & \\
2327.5+0323 & 1RXS J232735.2+03233&23 27 34.2 & ~~03 23 01.6 &   9.83$\times10^{-13}$ & unclassified & \\
2327.6+0838 & &23 27 40.1 & ~~08 38 30.4 &   1.73$\times10^{-13}$ & unclassified & \\
2328.4+0854 & 1WGA J2328.4+0852 &23 28 29.3 & ~~08 54 08.5 & 1.52$\times10^{-13}$ & unclassified & \\
2328.9+0854 & 1WGA J2328.9+0853 &23 28 54.6 & ~~08 54 07.8 &   4.68$\times10^{-13}$ & unclassified & \\
2335.0~-5601 & &23 35 00.7 & ~-56 01 15.0 &   3.09$\times10^{-13}$ & unclassified & \\
2335.2~-5609 & &23 35 12.1 & ~-56 09 08.9 &   1.80$\times10^{-13}$ & unclassified & \\
2359.9+0833 & 1RXS J235959.1+08335 &23 59 58.5 & ~~08 33 56.3 &   1.99$\times10^{-12}$ & unclassified  &\\
\hline
\end{tabular}
$^{(a)}$Zamorani et al 1999, $^{(b)}$ Veron \& Veron 1996, $^{(c)}$ Stocke \etal 
1991, 
$^{(d)}$ Vikhlinin \etal 1998, $^{(e)}$ Puchnarewicz \etal 1997,
$^{(f)}$Schmidt \etal 1998, $^{(g)}$Lehmann \etal 2000, 
$^{(h)}$ Fiore \etal 1999, $^{(i)}$ Burbidge 1995, $^{(j)}$ Marziani \etal 1996, 
$^{(k)}$ Boyle \etal 1991, $^{(l)}$ Hewett \etal 1995, $^{(m)}$ Ciliegi \etal 1995, 
$^{(n)}$ Kulkarni \etal 1997, $^{(o)}$ Gorosabel \etal 1998, 
$^{(p)}$ Hewitt \& Burbidge 1993, $^{(q)}$ Greiner \etal 1996, $^{(r)}$ Warren \etal 1991 
\end{minipage}
\end{table*}

\begin{figure*} 
\centering{
\epsfig{figure=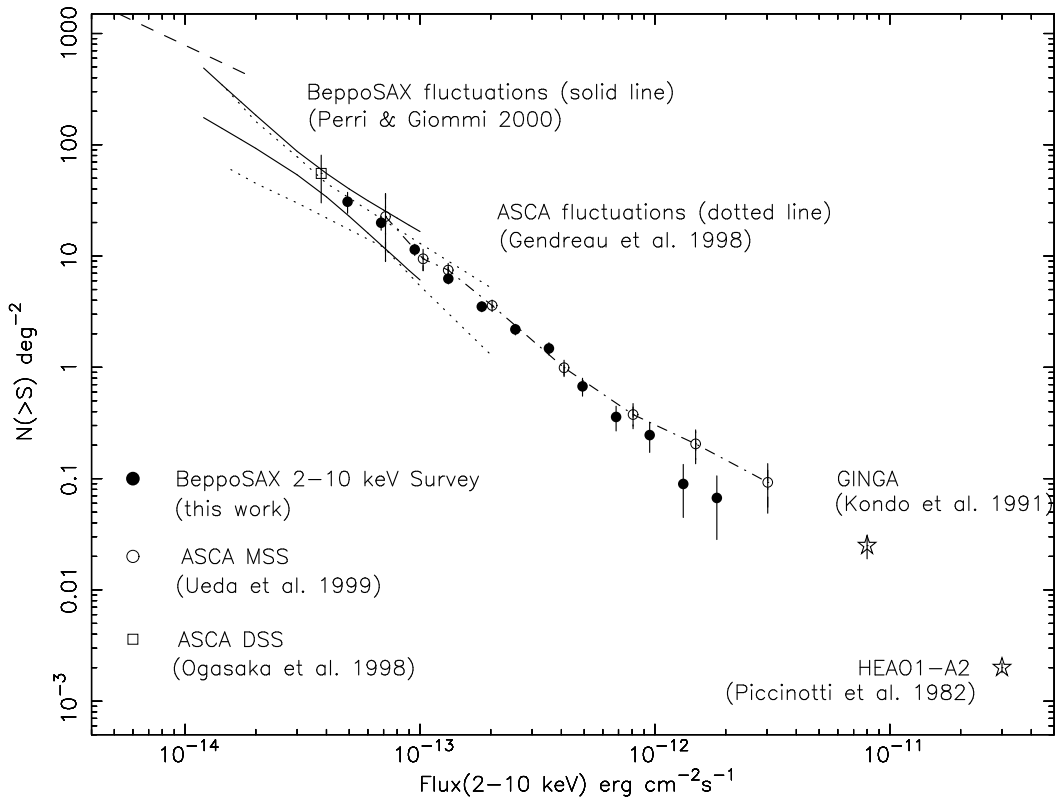,height=16.0cm,width=12.cm,angle=-90} }
\caption{The 2-10 keV cumulative logN-logS function and the 
(1 sigma) constraints from the CXB fluctuation analysis derived from 
\sax and ASCA data. 
Data points from the ASCA Deep Survey, the ASCA Medium Sensitivity 
Survey, Ginga and HEAO1-A2 are also shown. The dashed line at the 
top left 
indicates the limit obtained imposing that the integrated flux from the 
discrete sources cannot not exceed 100\% of the CXB.} 
\label{fig1} 
\end{figure*} 

\section{The logN-logS} 

The X-ray flux distribution of the 177 sources of table 1  
has been combined with the sky coverage of figure 1 to 
estimate the logN-logS function in the 2-10 keV band. 
The cumulative logN-logS, calculated assuming an average source 
spectral slope $\alpha$ of 0.7, is plotted in figure 2 (filled circles) 
together with the counts derived from ASCA (open circles) and from 
other satellites (open stars). The constraints obtained from the \sax 
CXB fluctuation analysis (see below) are also plotted in the usual ``bow 
tie'' shape.   
Table 2 gives the sky coverage in numerical form together with 
the cumulative source counts. 
The \sax logN-logS between $ 1 \times 10^{-12} $ and 
$5\times 10^{-14}$ \cgs is well described by the relation 
\begin{equation}
           N(>S)=10.7 \times (S/10^{-13})^{-1.65\pm 0.1}
\end{equation}	
where $N(>S)$ is the number of sources per square degree with flux 
larger than $S$ in the 2-10 keV band. The best fit values (and 
1 sigma error) for the normalization and slope of the logN-logS 
have been calculated by means of a maximum likelihood method (Murdoch, 
Crawford, and Jauncey 1973) and are in good agreement with those of
ASCA (Della Ceca \etal 1999b). 

The logN-logS slope is steeper than the "euclidean value" of 1.5, 
probably indicating that some amount of cosmological evolution 
is present.

The agreement with all the ASCA surveys, both in normalization and slope,
is very good as can be clearly seen in figure 2. 
The first results on the much deeper Chandra 2-10 keV logN-logS 
(Mushotsky \etal 2000) is also fully consistent with our data, although 
the statistics of the Chandra survey in the overlapping flux range is 
still very limited. 

The contribution of the logN-logS sources to the CXB at our flux limit is 
of the order of 25\% (using a CXB intensity of 
$2.3\times 10^{-11} erg~cm^{-2}~s^{-1}~deg^{-2}$ as estimated from 
\sax MECS data, Vecchi \etal 1999, Perri \& Giommi 2000).

\section{Simulations and the problem of source confusion} 

The size of the MECS PSF depends on energy in such a way that
the deep exposures in the 2-10 \sax survey are significantly more 
affected by source confusion then those of the harder (5-10 keV) 
HELLAS survey.
To properly address the effects of source confusion in the \sax 2-10 
keV survey we have carried out extensive simulations using the data 
simulator of the \sax SDC (Giommi \& Fiore 1997). A description of 
this tool can be found at the web page http://www.sdc.asi.it/simulator. 
                                   
One hundred MECS fields with exposures of 100,000 seconds each 
were generated and subsequently analyzed following the same 
procedure used for the survey. Each field included pointlike sources 
following  a logN-logS distribution equal to that measured in the real 
data above  $S=1\times 10^{-13}$\cgs and extended down to 
$S=1\times 10^{-14}$ \cgs. 

The analysis of these simulated fields resulted in the selection of  
a large sample of sources that was used to estimate the logN-logS 
parameters. 
 
Although some cases of source confusion were clearly present no 
significant bias in the estimation of the logN-logS slope or normalization 
could be found down to a flux of approximately $5\times10^{-14}$ \cgs. 
At lower flux levels a number of sources could still be detected but in  
this regime source confusion introduces severe biases in the 
determination  of source flux and accurate positions (see also the 
results of Hasinger \etal 1998). 

In the following we take $5\times10^{-14}$ \cgs as the confusion limit
of the MECS instrument.

\setcounter{table}{1}
\begin{table}
\begin{minipage}{80mm}
\begin{center}
\caption{\sax 2-10 keV Survey Sky Coverage and logN-logS data.}
\begin{tabular}{ccl}
\hline \hline
 2-10 keV Flux       & Area    & $N(>S)$            \\
   \cgs              & $deg^2$ & $deg^{-2}$         \\
\hline
$4.9\times10^{-14}$  &  0.18   & 30.8  $\pm$  6.6   \\
$6.3\times10^{-14}$  &  0.65   & 21.2  $\pm$  3.1   \\
$1.0\times10^{-13}$  &  2.9    & 10.3  $\pm$  1.2   \\
$1.7\times10^{-13}$  &  12.8   & 4.2   $\pm$  0.4   \\
$2.8\times10^{-13}$  &  29.6   & 1.98  $\pm$  0.22  \\
$3.5\times10^{-13}$  &  36.7   & 1.48  $\pm$  0.19  \\
$5.8\times10^{-13}$  &  44.3   & 0.51  $\pm$  0.11  \\
$7.4\times10^{-13}$  &  44.7   & 0.31  $\pm$  0.08  \\
$1.2\times10^{-12}$  &  44.7   & 0.089 $\pm$  0.047 \\
\hline
\end{tabular}
\\
\end{center}
\end{minipage}
\end{table}

\section{CXB fluctuation analysis} 
 
In order to extend our study of the logN-logS relationship  
beyond the MECS confusion limit of $5\times10^{-14}$ \cgs we have  
performed an analysis of the spatial fluctuations of the  
2-10 keV cosmic background. For reasons of brevity here we only 
describe the basic steps of the procedure leaving the details to a 
dedicated paper (Perri \& Giommi 2000).
 
We have used the set of 22 non-overlapping high galactic latitude 
MECS fields pointed at ``blank'' parts of the sky which had exposures 
ranging from 25,000 to 270,000 seconds.
To maximize the signal-to-noise ratio and to avoid complications 
introduced by the MECS window support structure, we have only 
considered the central 8 arcminutes in each image. Each circular region
was then divided in 4 equal quadrants for a total of 88 independent  
measurements of the CXB. 
 
Net counts were extracted between channel 44 and 200 (2.0-9.0 keV) 
and converted to flux in the 2-10 keV band assuming a power law 
spectrum with energy index $\alpha = 0.7$. 
The MECS internal background (about half of the total signal in the
central parts of the field of view) was estimated as in Vecchi et al. 
(1999) using 3.85 Ms of MECS data accumulated over a period of three 
years during intervals when the sky was occulted by the Earth.  
Special attention was used to take into account of instrumental 
effects such as slight time variations of the MECS internal background 
and the non-negligible size of the 2-10 keV PSF compared to the 
regions where the CXB signal was measured. 
 
The observed CXB flux distribution was compared to a number of 
analytically predicted distributions corresponding to different 
trial logN-logS parameters. A maximum likelihood test has been used to estimate the 
best fit and the 68\%, ($\Delta S=2.3$) constraints to the logN-logS 
slope and normalization which are plotted in figure 2 together with the 
ASCA and \sax  logN-logS. A very good agreement is found in the 
overlapping flux range, whereas the constraints are too weak to detect  
any slope change just below the ASCA and \sax logN-logS. 
The additional constraint imposed by the fact that the integrated flux 
from the logN-logS sources cannot exceed the observed CXB intensity 
(dashed curve at faint fluxes), however implies that the steep slope 
measured above $5\times 10^{-14}$ \cgs cannot extend much below 
$2-3 \times 10^{-14}$ \cgs.
Similar results at somewhat higher fluxes have been obtained with the 
analysis of the ASCA fluctuation analysis (figure 2 and Gendreau \etal 1998). 

Our findings are fully consistent with the first measurements of the 
2-10 keV logN-logS at faint fluxes using Chandra data (Mushotzky \etal 
2000).

\section{Hardness ratio analysis} 

Earlier \sax results (Giommi et al. 1998, Fiore \etal 2000a) and ASCA 
surveys (e.g. Ueda et al. 1999, Della Ceca et al. 1999a) convincingly 
showed that a substantial fraction of serendipitous  sources in the 
2-10 keV band are either very flat or show evidence for large intrinsic 
absorption. 
To test for the presence of hard/absorbed sources in our survey
we have carried out  a hardness ratio analysis following a procedure 
that is somewhat different than that of the HELLAS survey.

We have divided the MECS bandpass in three parts: a soft band S 
(1.3-2.5 keV), a medium band M (2.5-4.4keV) and a hard band H (4.4-9.6 keV). 
We have then defined the softness ratio SR=S/M which is sensitive to 
absorption from \nh $\sim 1\times 10^{21}$ up to $\sim10^{23} cm^{-
2}$, and a hardness ratio HR=H/M which is less affected by absorption 
and allows a better estimation of intrinsic spectral slopes. 

Converting the softness ratio into spectral slopes, setting \nh equal to 
the Galactic value, we see that there is a very wide range of spectral 
slopes.
In particular, out of the 177 sources in our survey 80 (45\%) have a 
energy slope flatter than 0.5, and about half of these flat sources 
have negative spectral slopes. This unusual spectral shape 
is probably not due to extreme spectra (never observed before), but 
rather the result of intrinsic absorption. In this hypothesis we can 
quantify the amount of intrinsic \nh inverting the softness ratios 
of these sources assuming a spectral index of 0.7. Figure 3 plots 
the distribution of the \nh in excess to the Galactic value, where 
it can be seen that intrinsic columns as high as several times 
$10^{22}~~cm^{- 2}$ are very common. 

\begin{figure} 
\epsfig{figure=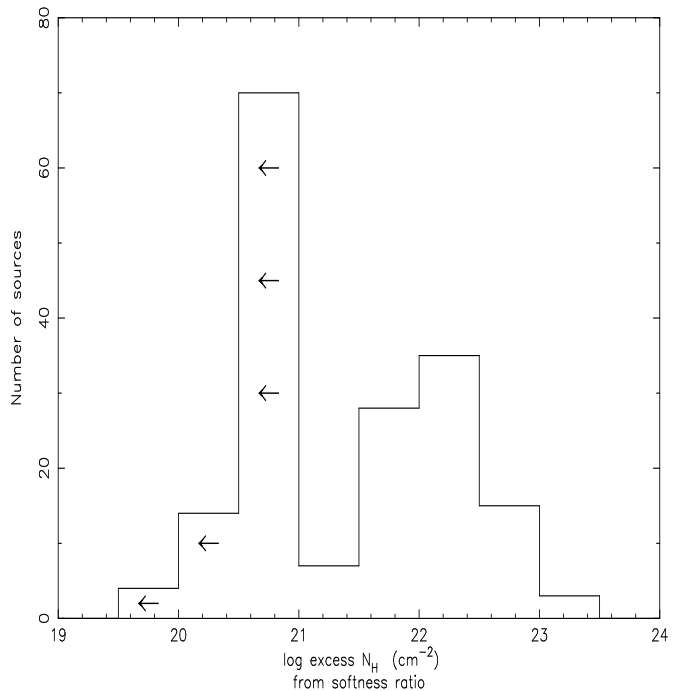,height=10.0cm,width=10.cm, angle=-90} 
\caption{The distribution of intrinsic \nh and upper limits, 
as derived from the softness ratio,
of the 177 X-ray sources of the \sax 2-10 keV Survey. A power law with 
energy index $\alpha = 0.7$ was assumed as the underlying spectral model.} 
\end{figure} 

\begin{figure}
\epsfig{figure=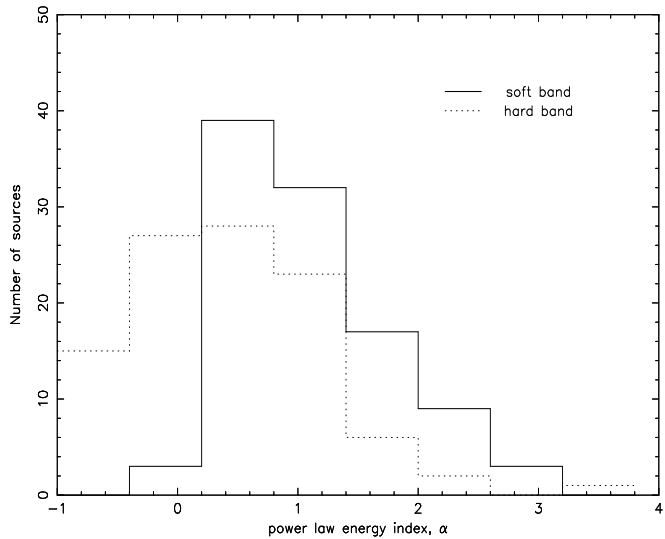,height=10.0cm,width=8.cm,angle=-90}
\caption{The distributions of spectral slopes in the soft (1.3-4.4 keV, 
solid histogram) and hard (2.5-9.6 keV, dotted histogram) band for the 
103 sources which are not affected by very large absorption and have 
been detected with a signal-to-noise ratio larger than 3.5. While there 
are no very flat slopes in the soft band (because they would have been
interpreted as evidence for absorption) a non-negligible fraction
of sources in the hard band are very flat ($\alpha_{hard} \lsim 0$). 
These spectra could arise from sources where strong absorption 
is present together with a soft component.}
\end{figure}

We have next considered the subsample of 103 sources detected with a signal 
to noise ratio (snr) level higher than 3.5 and for which the amount 
of intrinsic absorption is less than $10^{22} cm^{-2}$. We have then 
calculated the distribution of spectral slopes in the 1.3-4.4 keV band 
($\alpha_{soft}$) and in the 2.5-9.6 keV band ($\alpha_{hard}$) 
using the softness and hardness ratios respectively.
The resulting distributions are plotted in figure 4, which shows that 
the spectral slopes in the hard band appear to be significantly flatter 
than those in the soft band. The main difference between the two
distributions, however, is the presence of a substantial number of very flat 
slopes in the hard band. Since these sources do not show high intrinsic 
\nh in the soft band their spectra must be concave (i.e. $\alpha_{soft} > 
\alpha_{hard}$ ), as is also apparent from figure 5 which plots 
$\alpha_{hard}$ versus $\alpha_{soft} $ for the subsample of objects 
detected with a snr larger than 4. These concave spectra could 
arise from heavily absorbed sources with superposed a soft component. 
Objects of this type have been found also in ASCA data 
(Della Ceca \etal 2000). If we assume that all the sources flatter 
than $\alpha_{hard} = 0.1 $ are also absorbed the distributions of 
$\alpha_{soft} $ and $\alpha_{hard}$ of the remaining sources (i.e. those 
not showing evidence of absorption, with or without soft component) 
become very similar ($\langle\alpha_{hard}\rangle=0.85 \approx 
\langle\alpha_{soft}\rangle= 0.89$).

To further study the differences between absorbed/unabsorbed  and 
steep/flat sources we have divided our sample into a ``steep'' (HR$<1.11$ 
90 objects) and a ``flat'' (HR$>1.11$ , 87 objects) subsample, and into 
an ``unabsorbed'' (SR$>0.55$ corresponding to \nh $< 1\times 10^{22} cm^{-2}$, 
123 objects) and ``absorbed'' subsample (SR$<0.55$, 54 objects). 
We have then calculated the logN-logS for each subsample after excluding 
all objects identified with stars or clusters of galaxies. 
The logN-logS functions of the ``absorbed/unabsorbed'' subsample are 
shown in figure 6 where it can be seen that the logN-logS of 
``absorbed'' sources (open circles) is significantly steeper than that of 
the ``unabsorbed'' sources (filled circles). The logN-logS curves for the  
``steep'' and ``flat'' subsamples are instead parallel (fig 6, top-right) 
suggesting that the reported spectral differences between bright and faint 
samples (Ueda \etal 1999, Della Ceca \etal 1999a) are more probably 
due to a changing percentage of absorbed sources rather than to a 
change of the intrinsic spectral slope with flux. 

This interpretation also explains why the hardening at faint fluxes is 
not present in the HELLAS survey which was carried out in a band (5-
10 keV) where  the effects due to \nh are much less important.

\begin{figure}
\epsfig{figure=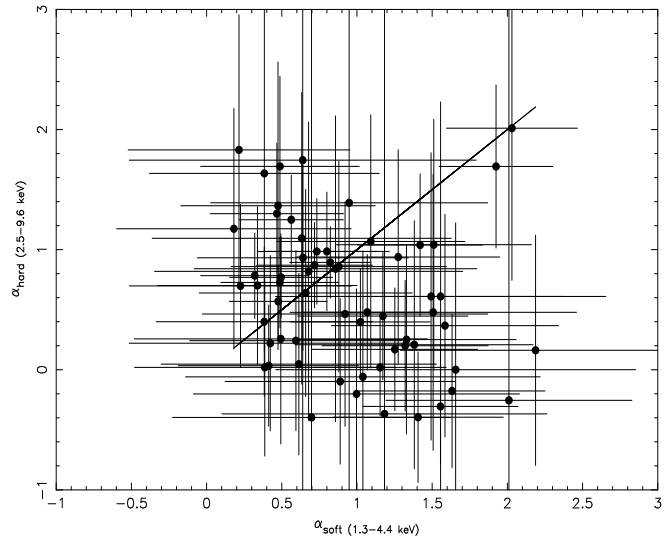,height=10.0cm,width=8.cm,angle=-90}
\caption{The power law energy index in the soft (1.3-4.4 keV) is plotted 
versus the (2.5-9.6 keV) slope for the 62 sources which are not affected 
by very large absorption and have been detected with a signal-to-noise 
ratio larger than 4.
The solid line marks the $\alpha_{soft}= \alpha_{hard}$ boundary.}
\label{fig13}
\end{figure} 

\begin{figure} 
\epsfig{figure=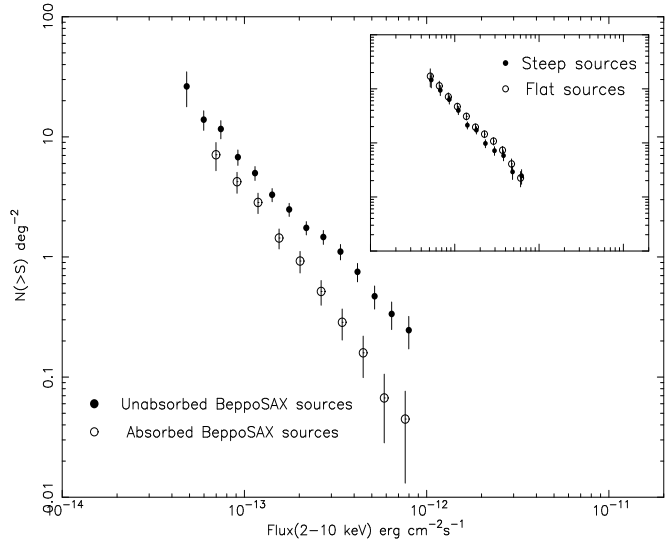,height=10.0cm,width=8.cm,angle=-90} 
\caption{The 2-10 keV logN-logS for the sub-samples of unabsorbed 
(filled circles) and absorbed sources (open circles) in the survey. Note 
that the slopes of the two logN-logS are significantly different; this 
could be the direct evidence of intrinsically different statistical 
properties but also the result of complex selection effects (see text). 
A similar comparison 
between the sub-samples of flat and steep spectral slope sources (as 
determined by the hardness ratio analysis in the 2.5-9.6 band) does 
not show any difference in the logN-logS slope (top-right panel).}
\label{fig11} 
\end{figure}

\section{Cross-identifications: Soft X-ray counterparts, Radio loud 
sources and Blazars candidates}

Cross-correlations with catalogs of known soft X-ray sources show that
many of the sources in our survey have a soft X-ray counterpart.  
The correlation with the \rosat WGA catalog (White, Giommi \& Angelini 
1994, Angelini \etal 2000) gives 58 matches within a correlation radius 
of 1.2 arcmin. 
We expect that the large majority of these matches are real since 
repeating the correlation under the same conditions after shifting the 
coordinates of our sources by a few arcminutes, the number of 
matches goes down to a very small number (1-3, see e.g. Giommi, 
Menna \& Padovani 1999 for an application of this technique). 
Increasing the correlation radius to 2 arcminutes the number of 
matches with and without the coordinates shift grows to 77, 
and to 12-14 respectively, suggesting that about half of the ROSAT/\sax 
sources between 1.2 and 2 arcminutes are real associations.  
We have also cross-correlated our catalog with the Rosat All Sky Survey
catalogs RASS-BSC, RASS-FSC (Voges \etal 1999, 2000).
Whenever a \sax source has a soft X-ray counterpart within 
2 arcminutes we report the \rosat (or other satellite) name in table 1; in 
case the distance is between 1.2 and 2.0 arcminutes the name of the 
soft X-ray source is followed by a question mark. 

To study the soft X-ray emission of our sources, whenever a \sax object 
is included in the field of view of one of the PSPC fields 
available from the \rosat public archive we have visually checked for 
possible soft X-ray detections. 
We have found that, with only one exception, all the \sax 2-10 keV 
sources have counterparts in the 0.2-2.0 keV band, if a reasonably 
deep image exist (i.e. exposure $> 2-3,000 s$) and the source is 
not under the PSPC window support structure.  
In the only instance where soft X-ray emission was not detected this 
could either be due to strong variability or to a genuine lack of soft 
X-ray emission in the cosmic source. Even if the latter is the correct
interpretation, we can safely conclude that the percentage of 2-10 keV 
sources that cannot be detected in the soft X-rays is less 
than $\sim 5$\%.

\setcounter{table}{2}
\begin{table}
\caption{Statistical identification of unidentified sources with radio 
counterparts.}
\begin{tabular}{lccccl}
\hline \hline
 Name & NVSS  & $m_R^{~~**}$  & \aox & \aro & Tentative \\
1SAXJ &flux&  &      &      & identification\\
 & (mJy) &  &      &      & \\
\hline
0243.1-0023  &  2.7   & 18.4       &  0.99&  0.35& HBL BL Lac  \\
0438.7-4727  & 130$^*$ & 21.7      &  0.65&  0.88& Blazar  \\
0916.3+2940  &  25.3  & 18.7       &  1.16&  0.51& Blazar  \\
1241.5+3251  &  2.8   & 18.4       &  1.1 &  0.3 & HBL BL Lac \\
1321.7-1635  &  3.6   & 19.8       &  1.01&  0.38& HBL BL Lac  \\
1635.5+5955  &  161   & 17.5       &  1.36&  0.55& Blazar  \\
2305.9+0902  &  3.4   & 16.8       &  1.47&  0.15& Blazar  \\
\hline
\end{tabular}
\\
$^*$4.85 MHz, from PMN catalog. \\
$^{**}$ Magnitude estimates are from the USNO catalog.
\end{table}

A cross correlation of our sample with the 
the NVSS catalog of radio sources (Condon \etal 1998), and the PMN 
catalog (Griffith \& Wright 1993) for sources south of 
dec = $-40^{\circ}$, resulted in 26 matches within a correlation 
radius =1.2 arcmin. 
Due to the very large number of faint NVSS sources, however, it is 
likely that a non-negligible fraction of these matches are accidental. 
By shifting the coordinates of our sources and re-running the 
correlation again we see that the number of spurious associations may 
be as high as 9-10. To reduce this number to a minimum we 
have only considered those matches where a) the precise NVSS 
position coincides with an optical counterpart on the Digitised Sky 
Survey (DSS) and b) the radio, optical and X-ray flux ratios are within 
the range seen in previous X-ray surveys.

Assuming that the (only) optical object within the NVSS error 
region of 1-5 arcsec is the correct counterpart of the \sax source 
we have derived the broad-band effective spectral indices \aox and 
\aro (calculated in the rest frame frequencies of 5 GHz, 5000 \AA $~$
and 1 keV) using the NVSS 1.4 GHz flux (extrapolated to 5.0 GHz), the 
magnitudes estimates of the USNO catalog (Monnet 1998) and the 
\sax fluxes.
We have then compared these values to those of known radio sources
to check if some of our objects fall in region of the \aoxaro plane 
that is typical of Blazars (Giommi, Menna \& Padovani 1999, Perlman 
\etal 1998). For 13 objects, seven of which previously 
unknown, this situation is indeed verified. We tentatively identify 
the latter sources as Blazars and list them in table 3 where column 1 gives 
the source name, column 2 the NVSS 1.4 GHz radio flux (or the PMN 
4.85 GHz  flux); column 3 the $m_R$ from USNO, columns 4 and 5 give 
the \aox and \aro, and column 6 gives a tentative identification based 
on the \aox and \aro values as in Giommi, Menna \& Padovani (1999). 

\section{The \vova statistics and cosmological evolution} 

The \vovm statistics (Schmidt 1968), and its extension \vova (Avni and 
Bachall 1980) to surveys with many flux limits like the \sax 2-10 keV 
survey, provides an effective and model-independent way to test 
for the presence of cosmological evolution. Due to the relativistic 
geometry of the Universe 
and to the K-correction, the \vova value depends on redshift, a 
quantity that is not known for the large majority of our sources.
We have nevertheless applied the \vova test (in the framework of a 
Friedmann cosmology with $q_0 = 0$) assuming a range of redshift 
values for all unidentified sources. 
Table 4 summarizes the results. Column 1 gives  
the assumed redshift, column 2 gives the \vovaave 
with its 1 sigma statistical uncertainty given by $\sqrt(12N)$,
where N is the number of sources.
For all values of the assumed redshift the \vovaave is significantly 
higher than 0.5, the expected value for a population of non-evolving  
objects. Even in the most conservative case where redshift of 
unidentified objects is fixed to 0.2, the \vovaave is about five sigma 
higher than 0.5 rejecting the hypothesis of a uniform distribution 
with a confidence higher than 99.99\% . This case is 
almost certainly too conservative, since the redshift distribution of the 
sources so far identified in the HELLAS and ASCA surveys  
reaches values well in excess 0.2 (Akiyama \etal 2000, La Franca \etal 
2000). 
If instead of using a single fixed redshift value we use 
Monte Carlo simulated redshifts, generated from the redshift 
distribution of the ASCA MSS survey (Akiyama \etal 2000), 
we find a mean value of \vovaave of 0.643 and a 90\% range 
of 0.637 and 0.650 in 100 simulation runs. We conclude that, despite 
the limited knowledge of the redshifts of our sources, the \vova test 
provides strong evidence for the presence of substantial cosmological evolution 
in the \sax 2-10 keV survey. 
\setcounter{table}{3}
\begin{table}
\begin{minipage}{80mm}
\begin{center}
\caption{\vova results, full survey (excluding stars and clusters 
of galaxies), 167 sources }
\begin{tabular}{cc}
\hline \hline
 Assumed redshift& \vovaave \\
 &  \\
\hline
  0.2 & 0.607$\pm$ 0.022 \\ 
  0.5 & 0.649$\pm$ 0.022 \\
  1.0 & 0.685$\pm$ 0.022 \\
MC z distribution $^{a}$ & 0.643$~\pm$ 0.022 $^{b}$ \\
              &  [0.637-0.650] $^{c}$  \\
\hline
\end{tabular}
\\
\end{center}
\end{minipage}

$^a$ Monte Carlo simulated redshifts based on the redshift distribution 
of the presently identified sources in the ASCA LSS survey. \\
$^b$ Average value of \vovaave with Monte Carlo simulated redshifts. \\
$^c$ 90\% range of \vovaave in 100 simulation runs.
\end{table}

Assuming a pure luminosity evolution law of the form 
$L(z)=L(z=0)\times(1+z)^C$ and the redshift distribution simulated as 
described above, we can quantify the amount of cosmological 
evolution present in our sample. We do that calculating \vova 
varying the value of the evolution parameter C until the \vovaave is 
0.5 (or 0.5 plus or minus the statistical error). This method gives $C= 
2.35\pm 0.2$, a value that is very close to those of $C=2.5-2.7$ 
derived in soft X-ray surveys (Maccacaro \etal 1991, Boyle \etal 1994, 
Page \etal 1996) and in combinations of ROSAT and ASCA data (Boyle 
\etal 1998). 
Given the limitations of our sample we do not attempt to fit 
more complex evolution laws (Miyaji \etal 2000).

As in section 6 we have also divided our sample in two parts: 
a) sources with high intrinsic absorption (\nh $> 1\times 10^{22}~cm^{-2}$ ) 
and b) sources with no evidence for high intrinsic absorption. The 
\vovaave for the two subsamples are listed in table 5 were it can be 
seen that in both cases they are significantly higher than 0.5 and that 
the value of \vovaave for the sample of absorbed sources is somewhat 
higher than that of the unabsorbed ones for all value of the assumed 
redshifts. The evolution parameter C, derived using the proper ASCA 
redshift distributions for absorbed and unabsorbed objects, (taken from 
Akiyama \etal 2000) is $C=3.0\pm0.4$ and $C=2.1\pm0.3$ 
respectively, indicating that a substantial amount of cosmological evolution 
is present in both subsamples and that this is possibly higher 
in absorbed sources. This different evolution rate in absorbed 
sources, however, could also be induced by complex selection effects since 
the conversion between count rate and flux (which depends on the amount 
of intrinsic absorption, the average luminosity and redshift) is most 
probably a function of intensity, rather than a constant value as we 
have assumed.

\setcounter{table}{4}
\begin{table}
\caption{\vova results, comparison between samples of absorbed and 
unabsorbed objects}
\begin{tabular}{lclc}
\hline \hline
 Sample& Number of & Assumed & \vovaave \\
             & objects       & redshift  &  \\
\hline
Absorbed & 51 & 0.2 & 0.656$\pm$ 0.040 \\
sources & 51 & 0.5 & 0.698$\pm$ 0.040 \\
& 51 & 1.0 & 0.734$\pm$ 0.040 \\
& 51 & MC $^{a}$& 0.656 $\pm$ 0.040 $^{b}$\\
&    &         & [0.645-0.666] $^{c}$ \\
Unabsorbed & 116 & 0.2 & 0.585$\pm$ 0.027 \\
 sources & 116 & 0.5 & 0.624$\pm$ 0.027 \\
 & 116 & 1.0 & 0.658$\pm$ 0.027 \\
 & 116 & MC $^{d}$& 0.629 $\pm$ 0.027 $^{b}$\\
 &  & & [0.620-0.636] $^{c}$ \\
\hline
\end{tabular}
\\
$^a$ Monte Carlo simulated redshifts based on the redshift distribution 
of the presently identified absorbed sources in the ASCA LSS survey. \\
$^b$ Average value of \vovaave with Monte Carlo simulated redshifts. \\
$^c$ 90\% range of \vovaave in 100 simulation runs. \\
$^d$ Monte Carlo simulated redshifts based on the redshift distribution 
of the presently identified unabsorbed sources in the ASCA LSS survey. \\

\end{table}

\section{Summary and Conclusions} 

We have selected a statistically well defined sample of 177 hard X-ray 
sources discovered in 140 \sax MECS fields covering 
$\sim 45$ square degrees of high galactic latitude sky. About 25\% of 
the  sources have been identified through cross-correlations with 
astronomical catalogs or using NED or the SIMBAD online systems;
96 sources have also been detected in soft X-ray images.

The 2-10 keV logN-logS in the flux range $5\times 10^{-14} - 
2\times 10^{-12}$ \ergs is steeper than that expected for a population 
of non-evolving sources in an euclidean universe. The best fit to
the cumulative distribution is $N(>S)=10.7\times (S/10^{-13})^{-1.65\pm 0.1}$ 
and is in good agreement with the counts derived from various ASCA surveys 
(Cagnoni \etal 1998, Ueda \etal 1998,1999, Della Ceca \etal 1999a). 

A CXB fluctuation analysis, performed on 22 fields centered on  
random parts of the sky, allowed us to constrain the logN-logS 
relationship down to about $1\times 10^{-14} $ \ergs.
The first estimate of the faint 2-10 keV Chandra logN-logS (Mushotzsky 
\etal 2000) is fully consistent with our results. 

The hardness ratio analysis reveals that a good fraction of 2-10 keV sources 
in the flux interval covered by our survey are intrinsically absorbed 
(figure 3). 
Figures 4 and 5 show that the range of spectral slopes is very wide  
and that some
objects are characterized by extremely flat (sometimes negative) slopes 
in the hard band but do not appear to be absorbed or flat in the soft band.
This result is in line with the findings of Giommi, Fiore \& Perri 1999 
who reported that the spectrum of extragalactic X-ray sources must include 
a soft component even in heavily cutoff objects, thus explaining why 
nearly all our 2-10 keV sources are also detected in soft X-ray 
images. 
Similar conclusions have been drawn from ASCA data by e.g. Della Ceca 
\etal (2000). One possibility is that these complex spectra could arise from
regions producing both an absorbed and an unabsorbed component
(``leaky absorber'') with the latter originating from 
scattering by warm material
above the absorbing torus, or from partial covering. Another possibility
is that the unabsorbed component comes from circumnuclear starburst 
regions or winds. 

Dividing our sources into different subsamples we showed that
the logN-logS of absorbed sources is steeper than that of unobscured 
objects. This trend is foreseen by CXB synthesis models 
and our findings are in broad agreement with the predictions of the  
models described in e.g. Comastri 1999.

A \vova test provides strong evidence for the presence of substantial 
cosmological evolution for any reasonable assumption on the redshift 
of our unidentified sources. Assuming Monte-Carlo simulated redshifts 
drawn from the observed distributions in the ASCA surveys, and
a luminosity evolution law of the form $L(z)=L(z=0)\times(1+z)^C$, we 
have been able to quantify the amount of cosmological evolution present 
in our survey. 
The evolution parameter C is estimated to be in the range C=2.1-2.5, a 
value that is similar to that found in soft X-ray surveys (Page \etal 1996, 
Boyle \etal 1994, 1998). Evolution is present in both 
unobscured (C=1.8-2.4) and absorbed sources (C=2.6-3.4) with the 
latter population possibly evolving faster.  
We note however that this difference could also be due to  
observational biases arising from possible correlations between 
physical or geometrical parameters such as luminosity, \nh etc.., 
or simply due to an intensity dependent count rate-flux conversion factor, 
rather than to a different rate of evolution of the central engine 
in absorbed and unabsorbed sources. Indeed CXB synthesis models
predict an increasing percentage of absorbed sources at faint fluxes
even assuming the same rate of cosmological evolution for all AGN (Comastri et
al 1999, Gilli \etal 1999).

Since cosmological evolution is one of the key ingredients in 
CXB models, confirmation (and a more precise assessment) of this 
result would have important implications for the understanding of 
the CXB.

We find that 13 objects are associated to radio-loud sources. A 
statistical identification technique, based on their location 
in the \aoxaro plane, allowed us to preliminarily identify 7 sources
with blazars, some of which are probably high energy peaked (HBL) BL Lacs. 
This number is 
consistent with the expectations from the logN-logS of 
BL Lacs in the soft X-rays (Wolter \etal 1991, Padovani \& Giommi 1995).

Nearly all the \sax 2-10 keV sources within the field of view of one of 
the \rosat images have a soft X-ray counterpart. This implies that 
if a population of 2-10 keV sources that are undetectable in the 
soft band exist, it must be a very small percentage of the total. 
Such a population was put forward as a possible 
explanation for the normalization of the 2-10 keV Ginga logN-logS (as 
derived from a fluctuation analysis) which was a factor 2-3 higher 
than the extrapolation of the 0.3-3.5 \einstein logN-logS assuming the 
canonical power law (energy) spectral slope of 0.7 (Butcher \etal 
1997). 
Our findings (see also Fiore \etal 2000a), and those of ASCA,
clearly show that the situation is more complex and rather different from 
the first simple interpretation of the CXB as the superposition of 
AGN with a ``canonical'' power law spectral slope of 0.7 which gave rise 
to the ``spectral paradox''.

CXB synthesis models went a significant step further incorporating 
into the picture the unified schemes for AGN and a number of parameters 
to describe the luminosity functions of absorbed and unabsorbed sources and 
their cosmological evolution.
The \sax and ASCA results are contributing to provide constraints 
to the parameters space and to reveal new phenomena. Chandra and XMM-Newton
will undoubtly significantly improve our understanding of the CXB. However,
the sensitivity has now gone past the point where different 
components (originating in the central engine, from partial covering, 
through reflections, and from starburst activity or from other 
circumnuclear sources) mix together causing severe complications to 
the interpretation of the data. A complete understanding of the 
CXB will probably have to wait for future  X-ray missions operating 
well above 10 keV providing for the first time an unobstructed view 
of the central engine of the sources making what was known as the 
diffuse background.


\section*{Acknowledgements} 

We thank F. Tamburelli for her crucial contribution to the 
upgrading of the XIMAGE detect routine, M. Capalbi for her help 
with the \sax archive, and A. Matteuzzi for his hard work on 
the MECS data reduction software.
We wish to thank R. Della Ceca for running the Brera Observatory 
Maximum Likelihood code on our logN-logS data and Y. Ueda for providing 
us with the AMSS logN-logS in numerical form. 
G.C. Perola, F. La Franca and A. Comastri are thanked for useful
discussion.

\noindent M. Perri acknowledges financial support from a Telespazio research 
fellowship.

\noindent Part of the software used in this work is based on the  
NASA/HEASARC FTOOLS and XANADU packages.    
This research has made use of the following on-line services: the 
ASI/BeppoSAX SDC Database and Archive System; the NASA/IPAC 
National Extragalactic Database, NED; the SIMBAD astronomical 
database, 
and the NASA Astrophysics Data System, ADS.



\end{document}